\documentclass[aps,prl,twocolumn,amsmath,amssymb,nofootinbib,superscriptaddress,floatfix,reprint,longbibliography]{revtex4-1}

\usepackage{graphicx}
\usepackage{dcolumn}
\usepackage{bm}
\usepackage{tabularx}
\usepackage{amsmath}
\usepackage{textcomp}
\usepackage{upgreek}
\usepackage{color}
\makeatletter

\newcommand{\Rmnum}[1]{\expandafter\@slowromancap\romannumeral #1@}
\makeatother

\usepackage[colorlinks=true, urlcolor=blue, linkcolor=blue, citecolor=blue, pdftex]{hyperref}
\usepackage{ulem} 
\usepackage[pdftex]{color} 

\begin{document}

\title{Superconductivity in layered Zintl phase LiSn$_2$As$_2$}

\author{Junjie Wang$^\dagger$}\affiliation{Beijing National Laboratory for Condensed Matter Physics, Institute of Physics, Chinese Academy of Sciences, Beijing 100190, China} \affiliation{School of Physical Sciences, University of Chinese Academy of Sciences, Beijing 100049, China}
\author{Xiangru Cui$^\dagger$}\affiliation{Beijing National Laboratory for Condensed Matter Physics, Institute of Physics, Chinese Academy of Sciences, Beijing 100190, China} \affiliation{School of Physical Sciences, University of Chinese Academy of Sciences, Beijing 100049, China}
\author{Yimin Wan$^\dagger$}\affiliation{State Key Laboratory of Surface Physics, Department of Physics, Fudan University, Shanghai 200438, China}
\author{Tianping Ying$^*$}\affiliation{Beijing National Laboratory for Condensed Matter Physics, Institute of Physics, Chinese Academy of Sciences, Beijing 100190, China}
\author{Shiyan Li}\affiliation{State Key Laboratory of Surface Physics, Department of Physics, Fudan University, Shanghai 200438, China}
\author{Jiangang Guo$^{*,}$}\affiliation{Beijing National Laboratory for Condensed Matter Physics, Institute of Physics, Chinese Academy of Sciences, Beijing 100190, China}\affiliation{School of Physical Sciences, University of Chinese Academy of Sciences, Beijing 100049, China}\affiliation{Songshan Lake Materials Laboratory, Dongguan 523808, China}

\date{\today}

\begin{abstract}
We report the superconductivity in the layered Zintl phase LiSn$_2$As$_2$, which is isostructural to NaSn$_2$As$_2$ and has a transition temperature ($T_{\mathrm{c}}$) of 1.6 K. Despite similar $T_{\mathrm{c}}$ and Debye temperatures, substituting of Na with Li considerably increases the upper critical field. Based on a systematically comparation of Sn$_4$As$_3$, NaSnAs, NaSn$_2$As$_2$,Na$_{1-x}$Sn$_2$P$_2$, SrSn$_2$As$_2$, and LiSn$_2$As$_2$, we propose that carrier doping, intimately related to the formation of lone-pair electrons, controls superconductivity in layered SnAs-based compounds rather than chemical pressure. The current findings provide a thorough and comprehensive understanding of Sn-based Zintl phase.

\end{abstract}

\maketitle

\section{\label{sec:level1}\expandafter{\romannumeral1}. introduction}
 The Zintl phase is the product where alkali metal or alkaline earth reacts with any post-transition metals (Group 13 or 14) or \textit{p} elements \cite{Zintl1,Zintl2}. There is usually a complete electron transfer from the more electropositive metal to anion clusters (Zintl ion), such as (Si$_4$)$^{4-}$ \cite{Si4}, (Tl$_4$)$^{8-}$ \cite{Tl4} and (As$_7$)$^{3-}$ \cite{As7}. When the Zintl ion is composed of infinite 2D layers, these compounds are called layered Zintl phase with numerous features including superconductivity \cite{ZintlSC1,ZintlSC2}, topological properties \cite{ZintlTOPO1, ZintlTOPO2, ZintlTOPO3, ZintlTOPO4} and thermoelectricity \cite{ZintlTHE,ZintlTHE1}. NaSn$_2$As$_2$, with a superconducting transition temperature at 1.6 K, has recently attracted a lot of attention \cite{ZintlSC1,NSA}. Later research discovered that NaSn$_2$As$_2$ can be prepared into few layers by mechanical and liquid-phase exfoliation. Substitution of As for a smaller P will increase the $T_{\mathrm{c}}$ to 2 K \cite{ZintlSC2}, while SrSn$_2$As$_2$ is proved to be a 3D Dirac semimetal without superconductivity \cite{SRSNAS1,SRSNAS2}. Detailed investigation suggests the $T_{\mathrm{c}}$ enhancement in Na$_{1-x}$Sn$_2$P$_2$ is due to additional charge donation from Na vacancies. However, the positive chemical pressure generated by P with a small radius can not be neglected. By the same token, replacing Na with a greater Sr will result in a negative pressure, which is consistent with the absence of superconductivity in SrSn$_2$As$_2$. As a result, the increase in $T_{\mathrm{c}}$ in the SnAs-based layered Zintl phase cannot be attributed solely to carrier doping or chemical pressure.
 
NaSnAs and NaSnP \cite{NaSnAs,NASNP}, on the other hand, are two closely NaSn$_2$As(P)$_2$ related compound by inserting one more Na ion in each SnAs(P) layer (Fig. \ref{fig1}(b)). Different from NaSn$_2$As$_2$ and Na$_{1-x}$Sn$_2$P$_2$ \cite{NSA,ZintlSC2}, NaSnAs and NaSnP are indirect band gap semiconductors with narrow gaps of about 0.31 and 0.62 eV, respectively \cite{ZintlTHE}. It is reported that NaSnAs(P) shows a much lower thermal conductivities than NaSn$_2$As(P)$_2$ due to the surplus lone-pair electrons. The lone pair is defined as a pair of valence electrons that are not shared with another atom to form a covalent bond. Sn and As(P) are covalently bonded to form a tetrahedron, as shown in Fig. \ref{fig1}(e). In the case of NaSnAs(P), with the one electron from Na, both Sn and As(P) enter a full outer shell structure, leaving two sets of lone-pair electrons dangling outside of the plane. SrSn$_2$As$_2$ follows the same logic, as Sr contributes two electrons to the system. Previous research found that surplus lone-pair electrons will generate strong inharmonic vibration and increase the Gruneisen parameters \cite{LONEPAIR1,LONEPAIR2,LONEPAIR3}. To fully understand the dominant factor of superconductivity in the series of Sn-based layered Zintl phase, new compounds that may generate chemical pressure without introducing extra carrier doping are highly desired.

In this work, we report the discovery of a new compound LiSn$_2$As$_2$ that is isostructural to NaSn$_2$As$_2$. Compared with NaSn$_2$As$_2$, LiSn$_2$As$_2$ has a similar $T_{\mathrm{c}}$ at 1.6 K, but a much-enhanced upper critical field ($H_{\mathrm{c}}$). First principle calculation reveals a significant change of the Fermi surface topology by the substitution of Li. Combined with experimental and theoretical analyses, we infer that the appearance of superconductivity in A-Sn-As system is dominated by the charge carrier doping, rather than chemical pressure. Finally, we provide a unified understanding towards the reported Sn-based layered Zintl phases, with NaSn$_2$Pn$_2$ serving as a parent phase for the series of superconductors.

\section{\label{sec:level1}\expandafter{\romannumeral2}. experiments}
 {\label{sec:level2}\expandafter \textbf{Synthesis}}
 
Stoichiometric amount of Li, Sn, and As were mixed and loaded into an alumina crucible, sealed into an evacuated quartz tube. Heated to 925 K with duration of 20 hours. Then, the sintered sample was taken out, reground, pelletized and heated to 900 K for 20 hours for adequate distribution. The sample preparation process are handled in a glove-box filled with argon gas. LiSn$_2$As$_2$ single crystals can be grown by using Sn-flux method, the sealed quartz tube was heated to 1200 K in 20 hours, slowly cooled to 1000 K with a rate of 3 K/h. The excess Sn is removed by centrifugation. Single crystal with size of 1.5 ×1.5 ×0.1 mm$^3$ can thus be obtained.

{\label{sec:level2}\expandafter \textbf{Characterization}}  

Powder X\text{-}ray diffraction (XRD) data were collected using a PANalytical X’Pert PRO diffractometer (Cu K$\alpha$-radiation) with a graphite monochromator in a reflection mode. Rietveld refinement of PXRD pattern was performed using Fullprof software suites \cite{FULLPORF}. The composition of Sn and As was determined by Energy Dispersive Spectroscopy (EDS), the elements ratio and element valence state information of Li, Sn and As were determined by inductively coupled plasma atomic emission spectroscopy and X-ray photoelectron spectroscopy (ICP-AES and XPS). Electrical resistivity and Hall resistivity were measured through standard six-wire method using the physical property measurement system (PPMS-9T, Quantum Design).

{\label{sec:level2}\expandafter \textbf{Theoretical calculation}}   

First\text{-}principles calculations were carried out with the density functional theory (DFT) implemented in the Vienna ab initio simulation package (VASP) \cite{VASP}. The generalized gradient approximation (GGA) in the form of Perdew\text{-}Burke\text{-}Ernzerhof (PBE) \cite{PBE} was adopted for the exchange\text{-}correlation potentials. We used the projector augmented\text{-}wave (PAW) \cite{PAW} pseudopotentials  with a plane wave energy of 800 eV; 1\textit{s}$^2$2\textit{s}$^1$, 4\textit{d}$^{10}$5\textit{s}$^2$5\textit{p}$^2$ and 4\textit{s}$^2$4\textit{p}$^3$ were treated as valence electrons for Li, Sn and As, respectively. A Monkhorst\text{-}Pack Brillouin zone sampling grid with a resolution 0.02 ×2$\pi$ \r{A}$^{\text{-}1}$ was applied \cite{PACK}. Atomic positions and lattice parameters were fully relaxed till all the forces on the ions were less than 10$^{\text{-}4}$ eV/\r{A}. The lattice parameters after relaxation were a = 4.03085 \r{A} and c = 25.85439 \r{A}, agreeing well with the experimental values within a deviation of 0.6$\%$. This confirmed the reliability of the calculation. Phonon spectra were calculated on a 2×2×1 supercell using the finite displacement method implemented in the PHONOPY code to determine the lattice dynamical stability of the structures \cite{CODE}. The band structure of NaSn$_2$As$_2$ was calculated for comparation using the same parameter except for the cut off energy was set to be 400 eV.
\section{\label{sec:level1}\expandafter{\romannumeral3}. results and discussion}
\begin{figure}[tp]
	\includegraphics[clip,width=8cm]{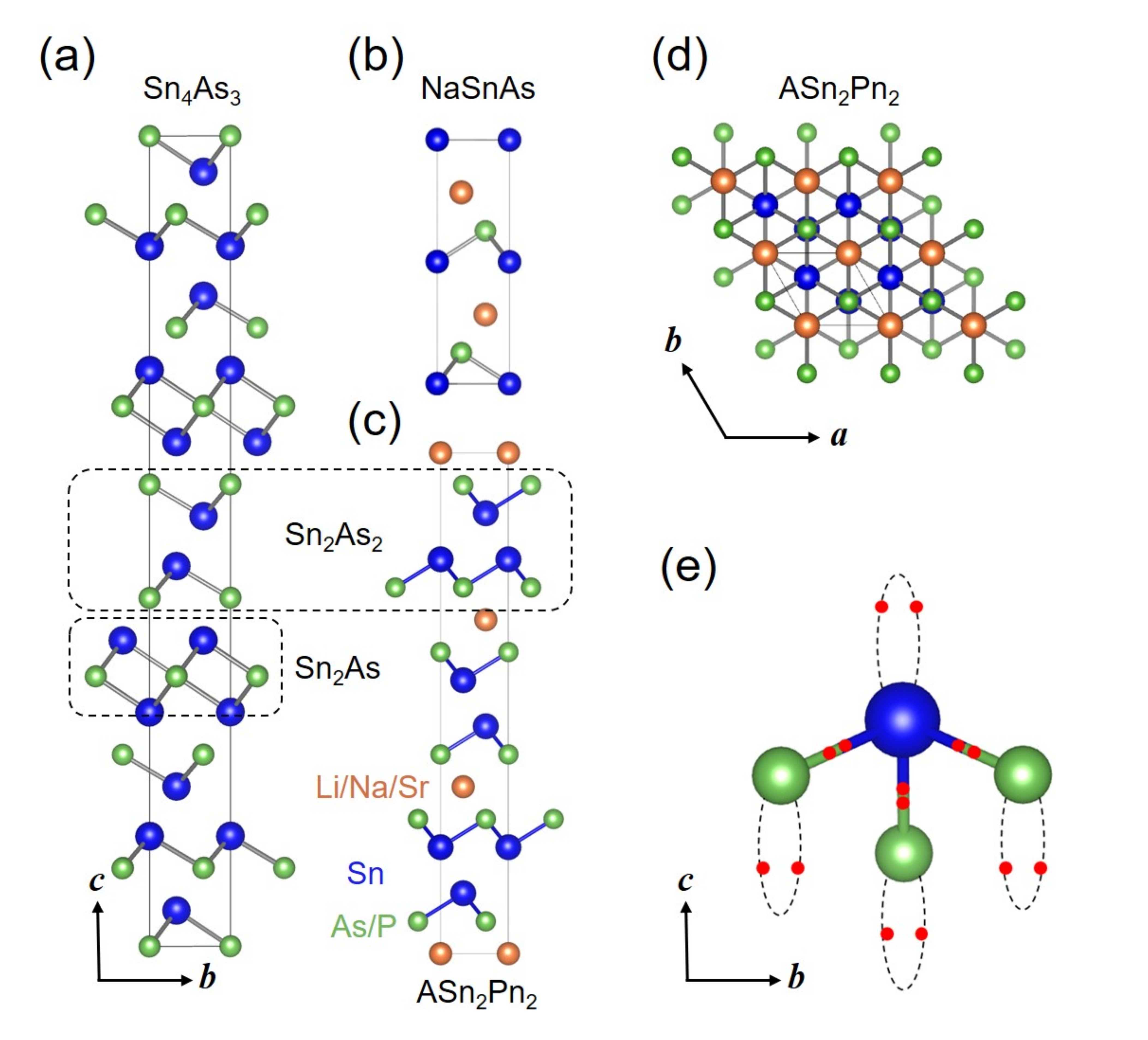}
	\caption{\label{fig1} (a-c) Crystal structures of Sn$_4$As$_3$, NaSnP, ASn$_2$Pn$_2$ (A = Li/Na/Sr, Pn = As/P). (d) The crystal structure of ASn$_2$Pn$_2$ along the c-direction. (e) Schematic diagram of long-pair electrons in Sn-based compound systems.}
\end{figure}

 LiSn$_2$As$_2$ is a new compound, isostructural to the well-studied NaSn$_2$As$_2$. Figure \ref{fig1} depicts all of the known Sn-based Zintl phase, including Sn$_4$As$_3$, NaSnAs and ASn$_2$Pn$_2$ (A= Li/Na/Sr, Pn = As/P), which crystallize in a space group of R\={3}m (No.166) and can be viewed as different stacking sequence of the SnPn layers with Na or other guest species. As shown in Fig. \ref{fig1}(a), Sn$_4$As$_3$ has two types of Sn-As configurations \cite{SN4AS3}. In one building block, each Sn (As) atom is coordinated by three As (Sn) atom to form a Sn$_2$As$_2$ layer, while the other Sn atom is joined to six As atoms to form a Sn$_2$As layer. As a result, Sn$_4$As$_3$ can be written as (Sn$_2$As)Sn$_2$As$_2$ (Fig. \ref{fig1}(a)), NaSnAs (Fig. \ref{fig1}(b)) and NaSn$_2$Pn$_2$ (Figs. \ref{fig1}(c) and (d)) share a similar stacking sequence of SnPn and A layers with the variation of Na in each SnPn layer or every two layers along the c axis. Figure \ref{fig1}(e) is the schematic diagram of the local electronic configuration of the SnAs tetrahedra. The covalent bonds between Sn and As are composed of the p orbitals, while the dangling lone-pair electrons are contributed by the \textit{s}$^2$ orbitals of Sn and As. For NaSnPn and SrSn$_2$As$_2$, the lone-pair orbitals are fully occupied, while the lone-pair on the Sn side is proved to be partially occupied in NaSn$_2$As$_2$.
 
\begin{figure}[tp]
	\includegraphics[clip,width=8.5cm]{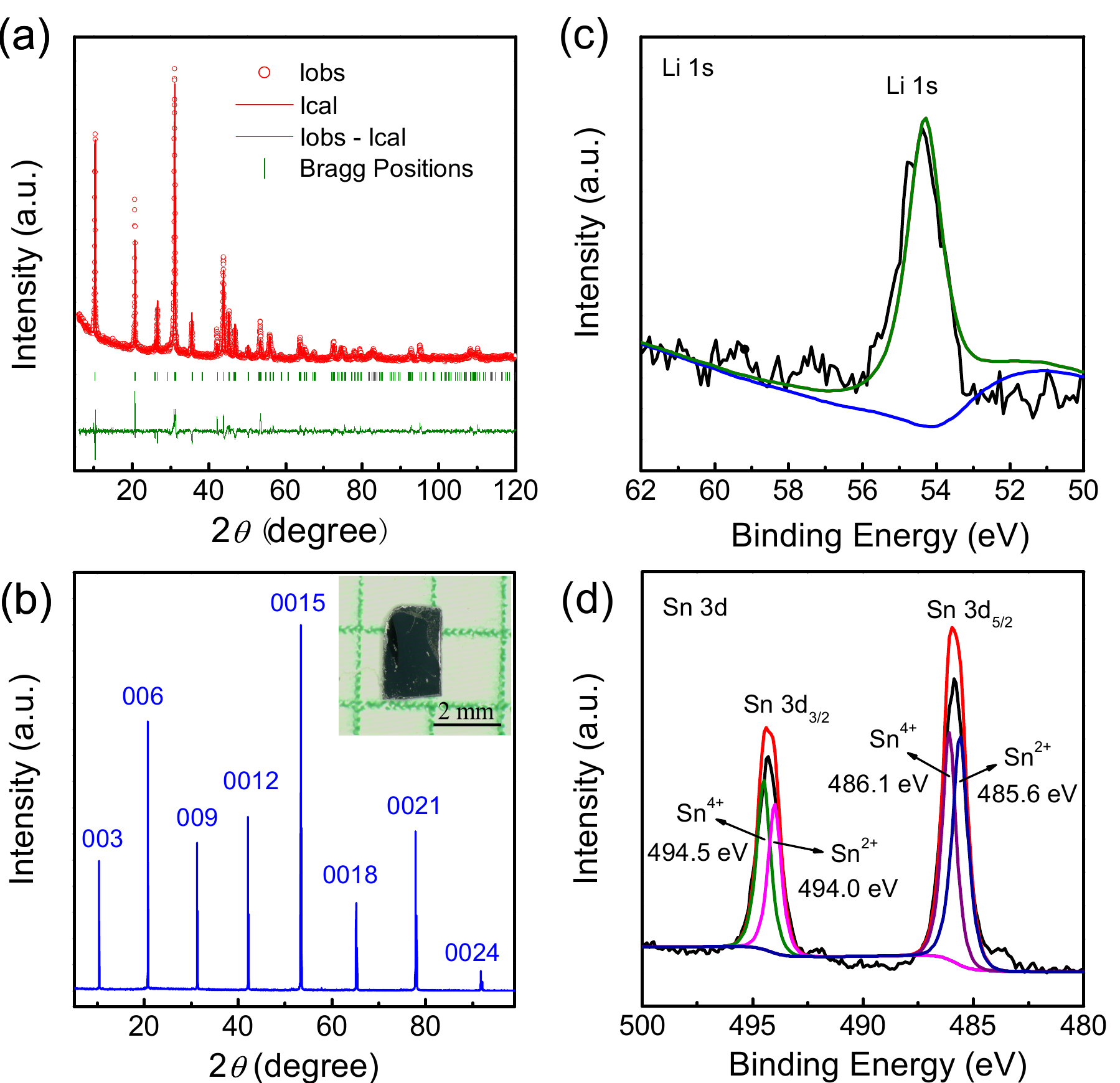}
	\caption{\label{fig2} (a) Rietveld refinements of powder x-ray diffraction (PXRD) pattern of LiSn$_2$As$_2$ at temperature at 300 K. (b)The XRD patterns of LiSn$_2$As$_2$ single crystal, where the preferred orientation of 00\textit{l} peaks can be seen. Inset shows the optical images of LiSn$_2$As$_2$ single crystal. (c, d) XPS spectrum of Li-1\textit{s} Sn-3\textit{d} orbit of LiSn$_2$As$_2$. }
\end{figure}
\begin{figure}[tp]
	\includegraphics[clip,width=8.5cm]{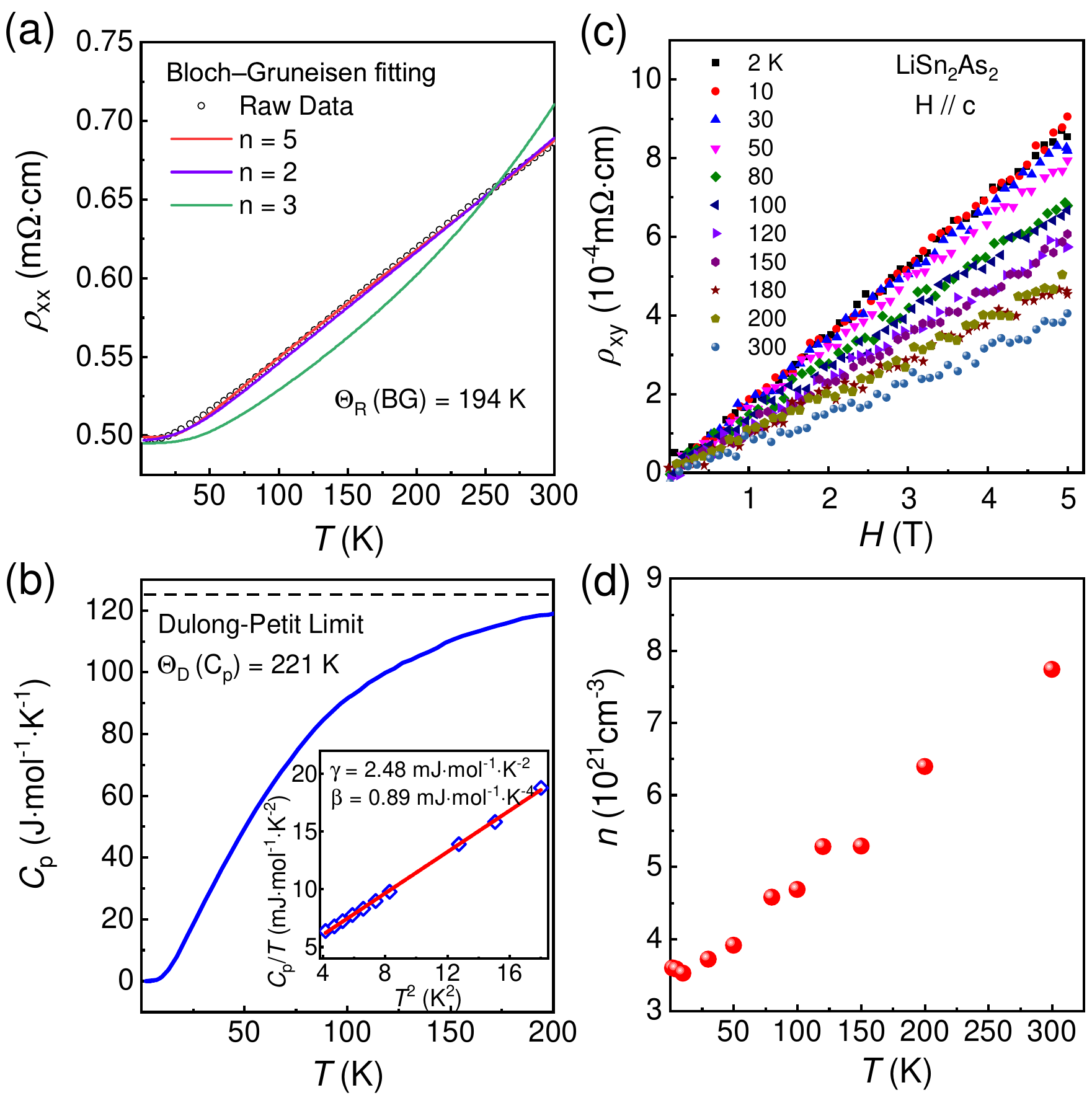}
	\caption{\label{fig3} (a) Resistivity of the LiSn$_2$As$_2$ single crystal from 2 to 300 K, the circles represent raw data and the solid lines are Bloch-Gruneisen fitting curve with \textit{z}=2,3 and 5. (b) Specific heat (C$_p$) as a function of temperature of LiSn$_2$As$_2$ single crystal. Inset shows C$_p$/\textit{T} as a function of \textit{T}$^2$ from 2-5 K. (c) Hall resistance of LiSn$_2$As$_2$ single crystal with \textit{H}//\textit{c} axis. (d) Carrier concentrations of LiSn$_2$As$_2$ single crystal as a function of temperature. }
\end{figure}

 Figure \ref{fig2}(a) shows the Rietveld refinements of powder diffraction pattern of LiSn$_2$As$_2$, yielding the lattice constants \textit{a} = 4.0072(3) Å and \textit{c} = 25.7008(6) Å. The refinement converged to the figures of merit \textit{R}$_p$=5.32$\%$, \textit{R}$_{wp}$=7.36$\%$ and $\chi^2$=2.63. The lattice change of the a and c axes are -0.02$\%$ and 6.82$\%$, respectively, as compared to NaSn$_2$As$_2$ with \textit{a}= 4.006\r{A}, \textit{c}= 27.581\r{A} \cite{ZintlTHE}, demonstrating the realization of chemical pressure. The diffraction pattern of a single crystal LiSn$_2$As$_2$ sample with 00\textit{l} preferred orientation is shown in Fig. \ref{fig2}(b), and the optical image is shown in the inset. XPS and ICP measurements were used to validate the element ratio and valence states in LiSn$_2$As$_2$. As the XPS spectra of Li 1\textit{s} orbital shown in Fig. \ref{fig2}(c), the binding energy of Li 1\textit{s} is 54.3 eV, consisting with that of Li$^+$. This shows that the Li ion has entirely lost its out shell electron to (Sn$_2$As$_2$)$^-$. Besides, the peaks of Sn 3\textit{d}$^{3/2}$ and 3\textit{d}$^{5/2}$ orbital locate at 494.3 eV and 485.9 eV, respectively (Fig. \ref{fig2}(d)), and both of them can be separated into Sn$^{4+}$and Sn$^{2+}$ peaks. Meanwhile, the chemical formula of stoichiometric Li:Sn:As = 1:2:2 is determined by ICP. LiSn$_2$As$_2$ single crystal exhibits a metallic behavior from 2 to 300 K (Fig. \ref{fig3}(a)). The residual resistivity ratio (RRR) value is calculated to be 1.36, which is lower than that of NaSn$_2$As$_2$ at 3.7 \cite{NSA}. To further comprehend its normal state behavior of LiSn$_2$As$_2$, we use the Bloch-Gruneisen (BG) model to fit the resistance curve \cite{BG2,BG3}
$$\rho(T)=\rho_{0}+4 R\left(\Theta_{R}\right)\left(\frac{T}{\Theta_{R}}\right)^{z} \int_{0}^{\frac{\Theta_{R}}{T}} \frac{x^{z} dx}{\left(e^{x}-1\right)\left(1-e^{-x}\right)}$$
 where \textit{T} stands for temperature and $\Theta_R$ for Debye temperature. For electron–electron interaction, \textit{s}-\textit{d} electron scattering, and electron scattering by phonons, the value \textit{z} could be 2, 3, or 5. The fitting curves of three alternative scattering mechanisms are shown in Figure \ref{fig3}(a). The raw data is best fitted by the curve with \textit{z}=5. In the LiSn$_2$As$_2$ system, the phonon dominates the scattering of normal state resistance. Other fitting parameters are $\rho_0$=0.499 m$\Omega$·cm, $\rho$($\Theta_R$)=0.614 m$\Omega$·cm and $\Theta_R$=194 K. To ensure that the BG fitting is correct, we measured the heat capacity of a single LiSn$_2$As$_2$ crystal, as shown in Fig. \ref{fig3}(b). The dotted line in Fig. \ref{fig3}(b) represents the Dulong-Petit limit, which is closed at C$_p$ (200 K)=118.64 J mol$^{-1}$ K$^{-1}$. The inset shows C$_p$/T as a function of T$^2$ in the temperature range of 2-5 K. The fitting formula is \cite{FT}
  
 \begin{figure*}[tp]
	\includegraphics[clip,width=17cm]{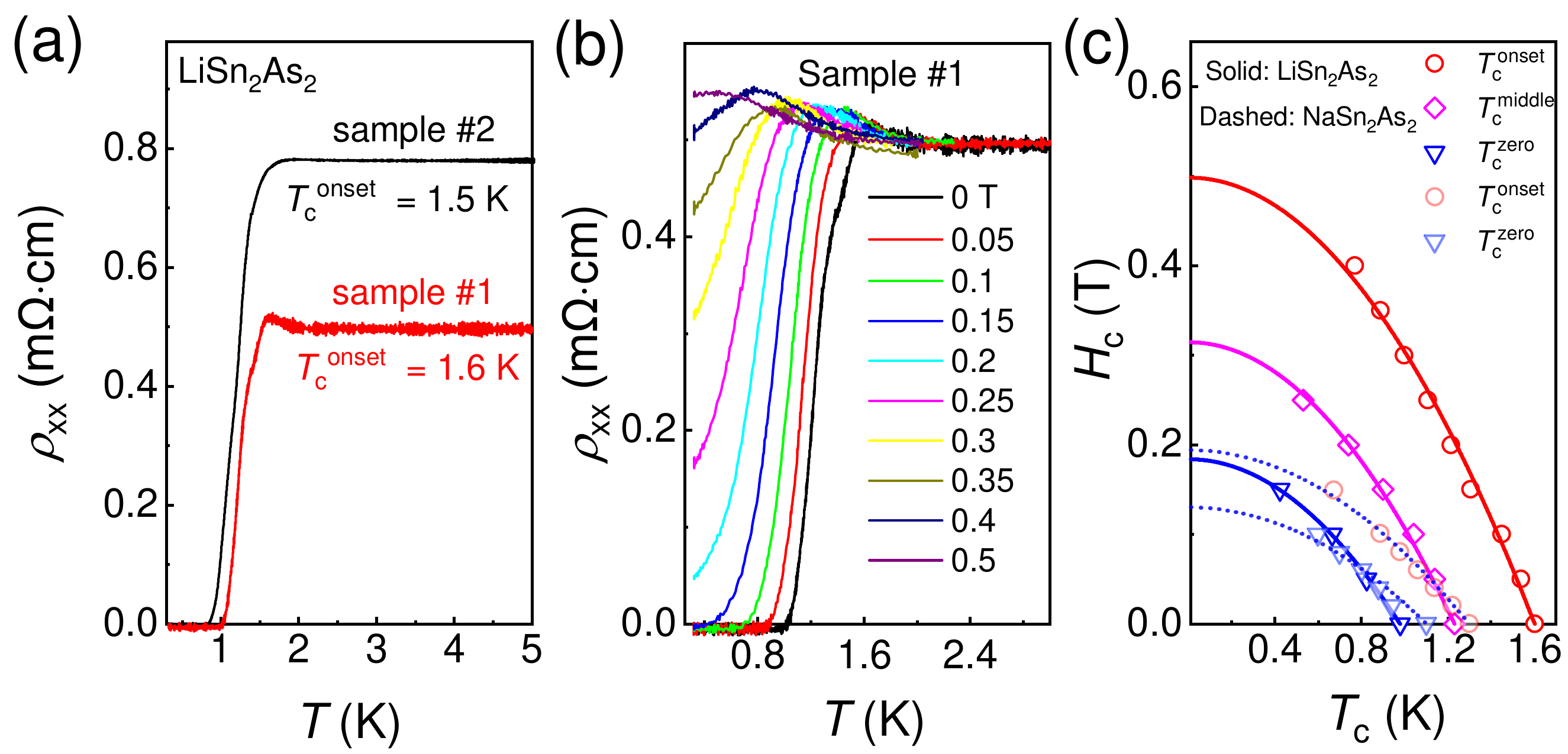}
	\caption{\label{fig4} (a) Low-temperature resistivity of two LiSn$_2$As$_2$ single crystals from 0.3 to 5 K. (b) Low-temperature resistivity at different magnetic fields parallel to the \textit{c} axis. (c) The upper critical fields ($H_{\mathrm{c}}$(0)) fitted by $T_{\mathrm{c}}^{onset}$, $T_{\mathrm{c}}^{middle}$ and $T_{\mathrm{c}}^{zero}$ with \textit{H}//\textit{c} axis. The dashed curve are the $H_{\mathrm{c2}}$(T) of NaSnAs$_2$ extracted from Ref \cite{ZintlSC1}}
\end{figure*}

 $$\frac{C_{P}}{T}=\gamma+\beta T^{2}$$
 $$\Theta_{D}=\sqrt[3]{\frac{12 \pi^{4} N_{A} r k_{B}}{5 \beta}}$$
 where $\gamma$ is the electronic specific heat coefficient and $\beta$ refers to the phonon specific heat coefficient. The fitted parameters are $\gamma$=2.48 mJ mol$^{-1}$ K$^{-2}$, $\beta$=0.89 mJ mol$^{-1}$ K$^{-2}$ and $\Theta_D$=221 K, which is close to the value obtained by BG model. The Debye temperature of LiSn$_2$As$_2$ is comparable to that of NaSn$_2$As$_2$ ($\Theta_D$=205K) and SrSn$_2$P$_2$ ($\Theta_D$=237K). The Hall resistance and extracted carrier density of LiSn$_2$As$_2$ are shown in Figs. \ref{fig3}(c) and (d). The positive slope and linear behavior of Hall resistance indicate that LiSn$_2$As$_2$ is a single band p-type metal. The carrier density (\textit{n}) is calculated by the single band model
 $${n}=\frac{1}{e R_{H}}$$
 where \textit{e} is elementary charge and R$_H$ refers to Hall coefficient. The carrier concentration of LiSn$_2$As$_2$ is 3.59×10$^{21}$ cm$^{-3}$ at 2 K and 7.74×10$^{21}$ cm$^{-3}$ at 300 K, confirming the metallic nature.

 Because LiSn$_2$As$_2$ and NaSn$_2$As$_2$ have similar transport behavior and Debye temperatures, we are interested in learning more about LiSn$_2$As$_2$'s ultra-low temperature behavior. The resistivity of two batches of LiSn$_2$As$_2$ single crystals in the temperature range of 0.3-5 K is shown in Fig.  \ref{fig4}(a). A sharp decline in resistivity can be seen at 1.6 K, with a zero resistance reaching at 1 K. There are slight differences between two batches of LiSn$_2$As$_2$ single crystals. The sample with lower residual resistance ($\#$2) shows a lightly lower $T_{\mathrm{c}}^{onset}$= 1.5 K than that of sample $\#$1, implying the small variation of $T_{\mathrm{c}}$ is caused by sample quality. The suppression of superconductivity at various external magnetic fields is seen in Fig. \ref{fig4}(b). In the presence of a magnetic field, the sample exhibits a tiny hump. This hump can also be seen in other instances, such as NdCeCuO and BiSrCaCuO \cite{NdCeCuO,BiSrCaCuO}, which is generally considered to be the result of imperfection of the single crystals. Figure \ref{fig4}(c) shows the upper critical fields $H_{\mathrm{c}}$(0) fitted by Ginzburg-landau equation
 $$H_{c}(T)=H_{c}(0)\left[1-\left(\frac{T}{T_{c}}\right)^{2}\right]$$
 The fitted $H_{\mathrm{c}}$(0) are 0.49 T, 0.31 T and 0.18 T for $T_{\mathrm{c}}^{onset}$, $T_{\mathrm{c}}^{middle}$ and $T_{\mathrm{c}}^{zero}$, respectively. We note that the acquired $H_{\mathrm{c}}$(0) of LiSn$_2$As$_2$ is much higher than that of NaSn$_2$As$_2$ (0.31 T for $T_{\mathrm{c}}^{onset}$ and and 0.15 T for $T_{\mathrm{c}}^{zero}$ , superimposed as dotted lines in Fig. \ref{fig4}(c)).

 \begin{figure*}[tp]
	\includegraphics[clip,width=18cm]{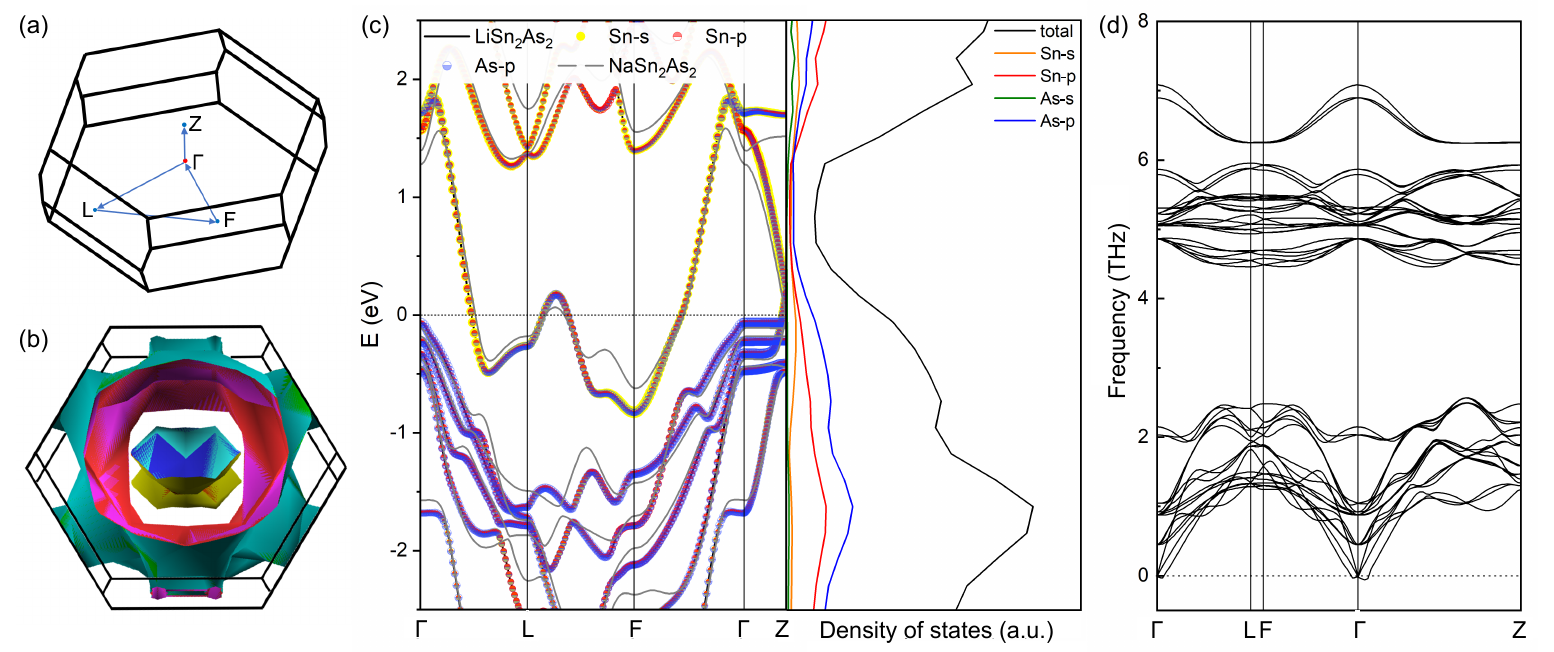}
	\caption{\label{fig5} (a) First Brillouin zone and the high symmetry K-path of LiSn$_2$As$_2$. The Fermi surface (b), band structure and partial density of states (PDOS) (c) and phonon spectrum (d) of LiSn$_2$As$_2$. The bands of NaSn$_2$As$_2$ are superimposed in Fig. 5(c) as gray curve for comparation. }
\end{figure*}

 Figure \ref{fig5}(a) depicts the initial Brillouin zone and the rhombohedral lattice's high symmetry route. The corresponding band structure and partial density of states (PDOS) are shown in Fig. \ref{fig5}(b). The bands of LiSn$_2$As$_2$ are more electron doped than the band structure of NaSn$_2$As$_2$ (shown as a dash curve), which is consistent with our Hall measurement. Furthermore, the Sn-\textit{s}, Sn-\textit{p}, and As-\textit{p} orbitals are mainly responsible for the bands around the fermi level (\textit{E}$\rm_F$). The \textit{s} orbital of Sn, in particular, contributes 18.3$\%$ of the total DOS at the \textit{E}$\rm_F$, whereas the \textit{s} orbital of As lies -10 eV below the \textit{E}$\rm_F$ (not shown here). LiSn$_2$As$_2$ has two sets of Fermi surface (shown in Fig. \ref{fig5}(c)), the cylindrical shape one is similar to that of NaSn$_2$As$_2$ \cite{AS2}, while the bowl shape one around Z point is attributed to the local hole-doped band characteristic of LiSn$_2$As$_2$, as the corresponding band in NaSn$_2$As$_2$ is located just below the Fermi level around Z point. The stability of the acquired LiSn$_2$As$_2$ is substantiated by the phonon calculations presented in Fig. \ref{fig5}(d). A neglectable imaginary frequency around $\Gamma$ is caused by numerical noise.
 
 Finally, we discuss the superconducting control parameter in the Sn-based layered Zintl phase. The volume contraction of 6.75$\%$ indicates that the iso-electron substitution of Na by Li produces an effective positive chemical pressure in NaSn$_2$As$_2$. The observation of superconductivity in LiSn$_2$As$_2$ and the comparable $T_{\mathrm{c}}$ rule out chemical pressure as a key driver in $T_{\mathrm{c}}$ increase. Meanwhile, NaSnAs and NaSnP are semiconductors with narrow band gaps at about 0.31 eV and 0.63 eV respectively, and SrSn$_2$As$_2$ is a semimetal \cite{NSA2,SRSNAS1}. This semiconducting or semimetallic behavior can be explained in terms of lone pair electrons, where the Sn-Pn tetrahedra have saturated electron pairs and the ionic bonds between the alkali metal Na and the Zintl ion SnPn do not have enough itinerant electrons, as briefly discussed in the introduction. Any hole doping can effectively tip the charge balance and boost conductivity in these double lone pair electrons, which serve as a charge reservoir. This is exactly the case of superconducting NaSn$_2$As$_2$, LiSn$_2$As$_2$ and Na$_{1-x}$Sn$_2$P$_2$ by partial removal of electrons from the lone pair electrons. This observation is consistent with our Hall results of \textit{p}-type charge carrier, and NaSnAs can be considered as a parent phase of the Sn-based layered Zintl phase. We infer that the (Sn$_2$As) slab in Sn$_4$As$_3$ provides less than one electron to the (SnAs) slab from this perspective. Because As or P have a stronger electronegativity than Sn, the charge transfer should take place on the Sn side, with the lone pair on the P side acting as a bridge of the charge transfer. Furthermore, the phonon vibration modes will be influenced by the quantity of lone pair electrons, which will affect the electorn-phonon interaction and alter $T_{\mathrm{c}}$. This could explain the slightly different $T_{\mathrm{c}}$ within NaSn$_2$As$_2$, Na$_{1-x}$Sn$_2$P$_2$ and LiSn$_2$As$_2$ (the slightly electron doping effect can be observed in Fig. \ref{fig5}(b)). Following this route, more exciting discovery can be anticipated. In SrSn$_2$As$_2$, for example, superconductivity may be achieved in a 3D Dirac semimetal by removing the electrons from the lone pair electrons through selective removal of Sr or substitution.
 
 \section{\label{sec:level1}\expandafter{\romannumeral4}. Conclusion}
 In conclusion, we report the crystal structure, transport properties, heat capacity and electronic structure of LiSn$_2$As$_2$. In the range of 2-300 K, LiSn$_2$As$_2$ exhibits metallic behavior. The $\Theta_D$ obtained by BG model and Debye model are 194 K and 221 K, respectively, which is closed to that of NaSn$_2$As$_2$. Ultra-low temperature transport measurements reveals that LiSn$_2$As$_2$ is a superconductor with $T_{\mathrm{c}}^{onset}$ at 1.6 K. The upper critical field of LiSn$_2$As$_2$ is $H_{\mathrm{c}}$=0.49 T, doubling that of NaSn$_2$As$_2$. We discovered that the amount of lone-pair electron acts as a charge reservoir and governs superconductivity by systematically analyzing the crystal structure and physical properties of Sn$_4$As$_3$, NaSnAs, NaSnP and ASn$_2$Pn$_2$. We anticipate similar phenomena can be realized in other layered Zintl phases.
 \section{\label{sec:level1}Acknowledgments}
 This work is financially supported by the National Key Research and Development Program of China (No. 2018YFE0202601 and 2017YFA0304700), the National Natural Science Foundation of China (No. 51922105 and 51772322), Beijing Natural Science Foundation (Grant No. Z200005).
 
\noindent
\\
$^\dagger$ These authors contribute equally\\
T.Y.: ying@iphy.ac.cn\\
J.G.: jgguo@iphy.ac.cn\\
\bibliography{LSA}
\end{document}